\documentclass[11pt]{article}

\usepackage[utf8]{inputenc}
\usepackage[T1]{fontenc}

\usepackage[margin=1in]{geometry}

\usepackage{amsmath,amssymb,amsfonts}
\usepackage{graphicx}
\usepackage{booktabs} 
\usepackage{textcomp} 
\usepackage{xcolor}  
\usepackage{tabularx} 
\usepackage{cite} 

\usepackage{algorithmic} 

\usepackage{hyperref}
\hypersetup{
    colorlinks=true,
    linkcolor=black,
    filecolor=black,      
    urlcolor=black,
    citecolor=black, 
    pdftitle={Case Study: Fine-tuning Small Language Models for Accurate and Private CWE Detection in Python Code},
    pdfpagemode=FullScreen,
    }

\usepackage{authblk}

\title{Case Study: Fine-tuning Small Language Models for Accurate and Private CWE Detection in Python Code}

\author[1]{Md. Azizul Hakim Bappy\thanks{0424312039@iict.buet.ac.bd}}
\author[1]{Hossen A Mustafa\thanks{hossen.mustafa@iict.buet.ac.bd}}
\author[1]{Prottoy Saha\thanks{prottoysaha@iict.buet.ac.bd}}
\author[2]{Rajinus Salehat\thanks{rajinussalehat@gmail.com}}

\affil[1]{Institute of Information and Communication Technology, Bangladesh University of Engineering Technology, Dhaka, Bangladesh}
\affil[2]{Hajee Mohammad Danesh Science and Technology University, Dinajpur, Bangladesh}

\date{} 

\begin{document}

\maketitle

\begin{abstract}
Large Language Models (LLMs) have demonstrated significant capabilities in understanding and analyzing code for security vulnerabilities, such as Common Weakness Enumerations (CWEs). However, their reliance on cloud infrastructure and substantial computational requirements pose challenges for analyzing sensitive or proprietary codebases due to privacy concerns and inference costs. This work explores the potential of Small Language Models (SLMs) as a viable alternative for accurate, on-premise vulnerability detection. We investigated whether a 350-million parameter pre-trained code model (codegen-mono) could be effectively fine-tuned to detect the MITRE Top 25 CWEs specifically within Python code. To facilitate this, we developed a targeted dataset of 500 examples using a semi-supervised approach involving LLM-driven synthetic data generation coupled with meticulous human review. Initial tests confirmed that the base codegen-mono model completely failed to identify CWEs in our samples. However, after applying instruction-following fine-tuning, the specialized SLM achieved remarkable performance on our test set, yielding approximately 99\% accuracy, 98.08\% precision, 100\% recall, and a 99.04\% F1-score. These results strongly suggest that fine-tuned SLMs can serve as highly accurate and efficient tools for CWE detection, offering a practical and privacy-preserving solution for integrating advanced security analysis directly into development workflows.
\end{abstract}

\vspace{1em} 

\section{Introduction}
Software security has become an undeniable cornerstone of our interconnected digital world. The increasing complexity and pervasiveness of software systems have, unfortunately, been paralleled by a rise in software vulnerabilities, making applications prime targets for malicious exploitation \cite{mittal2024software}. Among the critical classes of software weaknesses, Common Weakness Enumerations (CWEs) stand out as fundamental flaws in code that can lead to a cascade of security issues \cite{martin2008common}. The ability to detect and remediate CWEs early in the software development lifecycle is therefore paramount, offering significant cost savings by preventing costly breaches and reducing the effort required for late-stage fixes \cite{bagnato2009security}. Traditionally, approaches to CWE detection have relied on static analysis tools and manual code reviews \cite{nikolic2021analysis}. While these methods are valuable, they can be resource-intensive, prone to false positives and negatives, and may struggle with the nuanced understanding required to identify complex weakness patterns.
The emergence of Large Language Models (LLMs) has heralded a new era in code analysis and software engineering \cite{vieira2024engineering}. These models, with their remarkable ability to understand and generate code, have shown impressive capabilities in various security tasks, including vulnerability detection and CWE identification \cite{dozono2024large}. However, a significant hurdle remains for organizations handling sensitive or proprietary code, such as in finance, healthcare, or government sectors. These entities often face stringent data governance policies and security protocols that restrict or entirely prohibit the transmission of their codebase to external cloud services for analysis \cite{kazim2015survey}. This creates a critical gap: while LLM-powered security tools offer promising capabilities, their inherent cloud dependency makes them inaccessible for on-premise security analysis in many contexts, leaving organizations with confidential codebases with limited options for leveraging state-of-the-art language model technology for CWE detection.
In response to this challenge, this paper explores the potential of Small Language Models (SLMs) as a viable and effective solution for on-premise CWE detection. SLMs, with their reduced parameter count and computational footprint, offer several key advantages. Firstly, and most importantly in this context, they can be deployed directly within an organization's infrastructure, ensuring that sensitive code remains secure and under local control \cite{sehrawat2021edgecomputing}. Secondly, SLMs are significantly less computationally demanding than their larger counterparts, requiring fewer resources for both deployment and inference, potentially leading to faster analysis and lower operational costs \cite{recasens2024towards}. This makes them particularly attractive for environments with limited computational resources or where rapid analysis is critical.
In this work, we hypothesize that a carefully fine-tuned Small Language Model can achieve high accuracy in CWE detection within Python code, providing a practical and privacy-preserving on-premise security solution. To validate this hypothesis, we focus on fine-tuning the codegen-mono model (350M parameters) \cite{nijkamp2022codegen} using a novel, semi-supervised approach. The key contributions of this paper are as follows:
\begin{itemize}
    \item We demonstrate the successful instruction-following fine-tuning of codegen-mono for enhanced CWE detection in Python code, achieving performance comparable to or exceeding more resource-intensive methods.
    \item We introduce a semi-supervised dataset generation methodology, leveraging a reasoning-focused LLM (Gemini-2.0-flash-thinking-exp-01-21) and rigorous manual verification, to create a targeted dataset for security-specific fine-tuning, addressing the challenge of data scarcity in this domain.
    \item Our experiments showcase the capability of a relatively small model to achieve demonstrably high performance in CWE detection, reaching near-perfect accuracy, precision, recall, and F1-score, highlighting the potential of SLMs for resource-constrained environments.
    \item This research addresses the critical need for on-premise security solutions by providing a practical and effective approach to leveraging language model technology for CWE detection in environments where data confidentiality is paramount.
\end{itemize}
The remainder of this paper is structured as follows: Section 2 will delve into related work in CWE detection and the application of language models in software security. Section 3 will detail our methodology for dataset creation and model fine-tuning. Section 4 will present and analyze the experimental results, demonstrating the performance of our approach. Section 5 will discuss the implications, limitations, and potential future directions of this research. Finally, Section 6 will conclude the paper, summarizing our key findings and their contribution to the field of software security.

\section{Related Works}
Our research intersects with several areas within software engineering and artificial intelligence, primarily Static Analysis Security Testing (SAST), the application of Large Language Models (LLMs) to code security, the use of Small Language Models (SLMs) for code tasks, and synthetic data generation for software engineering.

\subsection{Traditional Tools}
Traditional Static Analysis Security Testing (SAST) tools form a baseline for Python code security, utilizing methods like Abstract Syntax Tree (AST) parsing (e.g., Bandit \cite{white2013bandit}), flexible pattern matching (e.g., Semgrep \cite{bennett2024semgrep}), and data/control flow analysis \cite{jeronimo2024techniques}. Beyond static analysis, other techniques contribute, including dynamic analysis frameworks like DynaPyt for runtime checks \cite{Eghbali2022DynaPytAD} and specialized machine learning models like BiLSTMs trained for vulnerability detection \cite{Farasat2024}. However, traditional SAST approaches, in particular, often face challenges with high false positive/negative rates \cite{charoenwet2024empiricalstudystaticanalysis} and require continuous, expert-driven maintenance of explicit rules to keep pace with evolving threats. Our work contrasts with these methods by leveraging the emergent pattern-recognition capabilities of a language model fine-tuned on specific vulnerability examples, aiming for high accuracy without reliance on predefined, manually curated rules.

\subsection{Large Language Models for Code Security}
The impressive performance of LLMs, such as OpenAI's Codex \cite{kumar2023open}, Google's PaLM variants \cite{maddy2024integrating}, and GPT-4 \cite{achiam2023gpt}, has spurred significant interest in their application to software security. Research has shown their potential in tasks like automatically detecting vulnerabilities from code descriptions or raw snippets \cite{Mahyari2024,du2024}, suggesting fixes for identified issues \cite{islam2024llm}, and even generating security test cases \cite{zhang2023well}. These studies often highlight the models' ability to understand code context and semantics better than traditional methods. However, as noted earlier, these powerful models are typically large (billions or trillions of parameters), computationally expensive, and often accessed via APIs, posing practical barriers related to cost, latency, and data privacy for security scanning of proprietary code. Our work specifically addresses these limitations by exploring the capabilities of significantly smaller models.

\subsection{Small Language Models for Code Tasks}
While LLMs grab headlines, there is a growing body of work focusing on Small Language Models (SLMs) models typically under 1 billion parameters tailored for code. Models like CodeGen \cite{nijkamp2022codegen}, CodeT5 \cite{wang2021codet5}, and smaller variants of StarCoder \cite{li2023starcoder} have demonstrated competence in tasks such as code completion, code summarization, and code translation. These models offer advantages in terms of deployment feasibility and reduced computational cost. However, their application to fine-grained security vulnerability detection, particularly through targeted fine-tuning for specific CWEs, has been less explored compared to their larger counterparts. Our research directly investigates this gap, assessing the extent to Gwhich a pre-trained SLM can be specialized for high-accuracy vulnerability detection post-fine-tuning.

\subsection{Synthetic Data Generation for Code and Security}
The performance of data-driven models heavily relies on the quality and quantity of training data. In specialized domains like software security, obtaining large, labelled datasets of real-world vulnerabilities can be challenging. Consequently, researchers have explored using generative models, including LLMs, to create synthetic data. Efforts exist in generating code for general software engineering tasks \cite{murphy2024combining}, augmenting existing datasets \cite{ugare2024syncode}, and specifically generating examples for security training or testing \cite{leinonen2024}. Our approach aligns with this trend by using an LLM (Gemini-Flash-Exp) to generate paired vulnerable and fixed code snippets. We contribute a specific methodology focused on generating data for targeted CWEs, emphasizing the role of a reasoning-focused generator model and subsequent human validation to ensure data quality for fine-tuning a security-focused SLM.

\begin{figure}[htbp] 
    \centering
    \includegraphics[width=\textwidth]{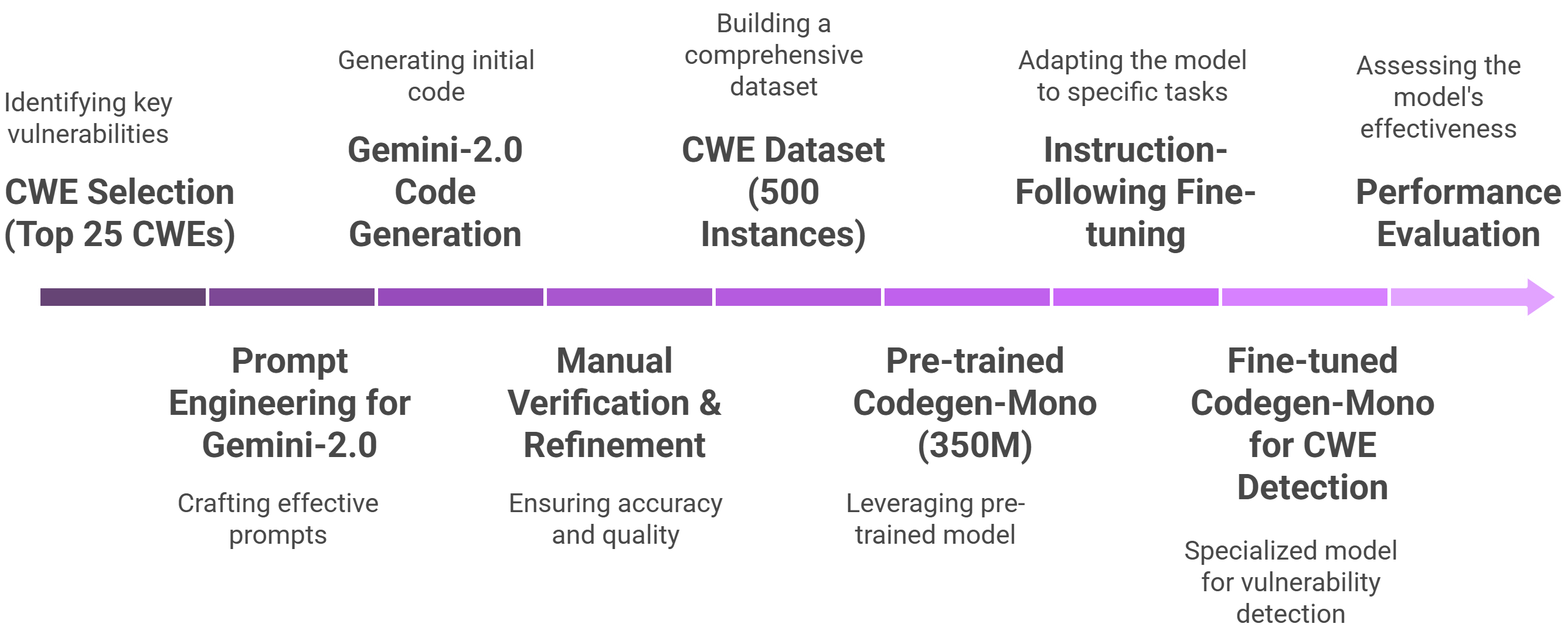} 
    \caption{Semi-Supervised Dataset Creation and Fine-Tuning Pipeline}
    \label{fig:pipeline}
\end{figure}

In summary, while extensive research exists on traditional static analysis, machine learning, and large language models for CWE detection, our work uniquely explores fine-tuning small language models for this task. This approach offers a novel, privacy-preserving, and computationally efficient solution, addressing a critical gap in the literature, particularly concerning on-premise deployment for confidential codebases and achieving remarkable performance with a significantly smaller model.

\section{Methodology}
Our approach focuses on fine-tuning a Small Language Model (SLM) to specialize in detecting specific Common Weakness Enumerations (CWEs) within Python code snippets. The methodology encompasses target selection, dataset creation, model selection, and the fine-tuning strategy. The whole process can be visualized in figure \ref{fig:pipeline}.
\subsection{Target Vulnerabilities (CWE Selection)}
To ensure practical relevance and focus on high-impact issues, we selected the MITRE Top 25 Most Dangerous Software Weaknesses list \cite{mitreMostDangerous} as the target for our detection model. This list represents common and critical vulnerabilities encountered in real-world systems, making it a suitable starting point for evaluating the feasibility of SLM-based detection.
\subsection{Dataset Curation: Semi-Supervised Synthetic Data Generation}
Creating a sufficiently large and accurately labelled dataset for vulnerability detection is often a bottleneck. To address this, we employed a semi-supervised approach combining automated generation with human oversight:
\begin{itemize}
    \item Generation Tool: We utilized Google's gemini-2.0-flash-thinking-exp-01-21 model via its API . This model was chosen for its reported reasoning capabilities, which we deemed beneficial for generating plausible code exhibiting specific logical flaws corresponding to CWEs.
    \item Generation Process: For each of the 25 selected CWEs, we iteratively refined prompts to instruct the Gemini model to generate: (a) Five distinct Python code snippets, each realistically demonstrating the specific CWE vulnerability. (b) For each vulnerable snippet, a corresponding "counter-example" snippet where the underlying issue causing the CWE was addressed and fixed.
    \item Prompt Engineering: Significant effort was invested in prompt engineering to guide the generator model towards producing code that reflects plausible real-world programming patterns, rather than trivial or overly simplistic examples.
    \item Human Review: Crucially, all generated code snippets (both vulnerable and secure) were manually reviewed by the authors to verify: (a) The correctness of the vulnerability classification (i.e., the vulnerable code actually exhibits the intended CWE). (b) The validity of the fix (i.e., the counter-example correctly addresses the vulnerability without introducing others). (c) The overall realism and relevance of the code examples. This human validation step was essential to ensure the quality and reliability of the training data.
    \item Final Dataset: This process resulted in a dataset of 500 labelled instances (25 CWEs × 10 examples × [1 vulnerable + 1 fixed]). Each instance was formatted according to a standard instruction-following structure: Instruction, Input, Output. 
\end{itemize}

\begin{figure}[htbp] 
    \centering
    \includegraphics[width=\textwidth]{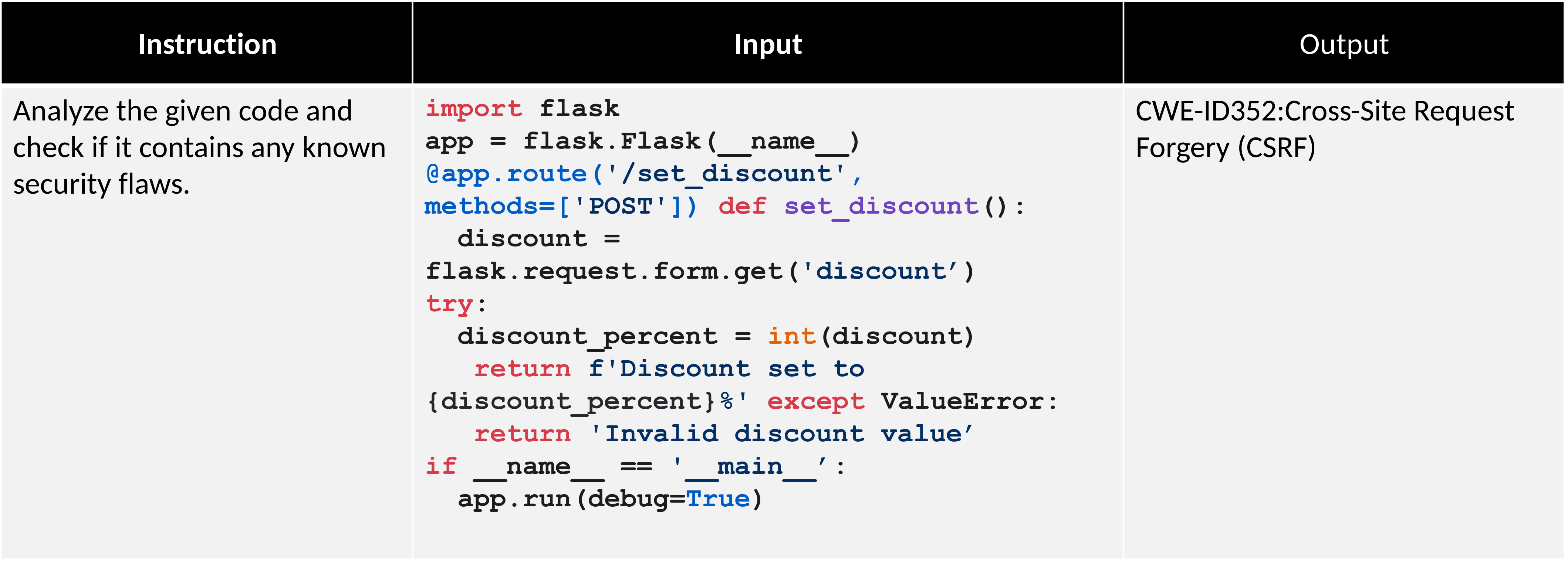} 
    \caption{Dataset Example}
    \label{fig:dataset}
\end{figure}

\begin{table}[htbp]  
\renewcommand{\arraystretch}{1.4}
\caption{Summary of Prior Works on Vulnerability Detection (ML/LLM Focus)}
\label{tab:summary}
\centering
\begin{tabularx}{\textwidth}{@{} l p{0.2\textwidth} >{\raggedright\arraybackslash}X >{\raggedright\arraybackslash}X @{}} 
\toprule
\textbf{Reference} & \textbf{Technique Used} & \textbf{Reported Performance} & \textbf{Remarks} \\
\midrule

Farasat \& Posegga \cite{Farasat2024} & ML (BiLSTM) & Python: Avg Acc=98.6\%, F1=94.7\%, Prec=96.2\%, Rec=93.3\%, ROC=99.3\%. & High performance on specific Python vulnerability detection task using ML. \\

Bagheri et al. \cite{bagheri2024towards} & Hybrid ML (Self-attention + CNN - Conformer) & F1 score of 99\% reported & High F1 reported, potentially domain-specific. \\

Singh et al. \cite{singh2024temporal} & ML (Logistic Regression) & CWE Prediction: Acc=0.66, Prec=0.65, Rec=0.66, F1=0.64. & Moderate performance using simpler ML for CWE prediction. \\

\midrule
Mechri et al. \cite{mechri2025secureqwen} & LLM Prompting (Chain-of-Thought) & Qualitative increase in F1 mentioned for CoT prompts on real vulns; no specific Python metrics. & Highlights prompt engineering benefits, lacks quantitative Python data. \\

Dozono et al. \cite{dozono2024large} & LLM Prompting (Various strategies) & Python F1: GPT-4o=0.80, GPT-4T=0.76, Gemini 1.5 Pro=0.75, CodeLlama-7b=0.72, GPT-3.5T=0.70, CodeLlama-13b=0.35. & Compares various LLMs for Python, shows GPT-4 variants leading. \\

Steenhoek et al. \cite{steenhoek2024err} & LLM Prompting (zero/n-shot, CoT variants) & Evaluated on C/C++: Low Balanced Accuracy (54.5\%); Python N/A. & Concludes current LLMs perform poorly; common prompting strategies ineffective (on C/C++). \\
\midrule

Shestov et al. \cite{shestov2025finetuning} & LLM Fine Tuning for JAVA CWE detection & Best F1 @ .86 for binary classification using WizardCoder & N/A \\ 
Li et al. \cite{li2024exploratory} & LoRa and IA3 Fine tuning approach for LLMs & 5-6\% improvement over base model & 9 CWE was tested using a 2.7b model, codegen. \\
Jiang et al. \cite{jiang2024investigating} & LLM Fine Tuning using LoRa approach & Best F1 achieved using Llama 2-7b model @ 87\% & N/A \\ 
\emph{This Work} & Instruction Following Fine Tuning, 350m parameters model & Accuracy: 99\%, Precision: 98.08\%, Recall: 100\%, F1 score: 99.04\% & High performance on Python CWEs with SLM fine-tuning. \\ 
\bottomrule
\end{tabularx}
\end{table}

\subsection{Model Selection}
We selected the codegen-mono 350M model \cite{nijkamp2022codegen} as our base SLM. This model was chosen because: (a) It is a publicly available, pre-trained model specifically designed for code-related tasks. (b) Its 350 million parameter size places it firmly in the "small" language model category, making it suitable for exploring feasibility for on-premise deployment and efficient fine-tuning. (c) Its focus on code generation/understanding provides a relevant foundation for the downstream task of vulnerability detection.

\subsection{Fine-tuning Strategy: Instruction Following}
We employed an Instruction-Following Fine-tuning approach to adapt the pre-trained codegen-mono model to our specific CWE detection task.
\begin{itemize}
    \item Data Format: The dataset was structured with three fields per instance: \emph{Instruction:} A directive telling the model what task to perform. \emph{Input:} The Python code snippet to be analyzed. \emph{Output:} The expected label (e.g., "Vulnerable - CWE-XXX" or "Secure"). An example of data instance is presented in figure \ref{fig:dataset}.
    \item Consistent Instruction: Through experimentation with various instruction phrasings and strategies (including varying instructions per row), we found that using a single, consistent instruction across all training examples yielded the best performance. This same instruction was subsequently used during the evaluation/inference phase. In our case we kept the instruction depicted in figure \ref{fig:dataset} for all the rows as well as during interface.
    \item Training Process: The model was fine-tuned using this structured dataset. We iterated through different hyperparameters (learning rate, batch size, epochs) and instruction formats to optimize performance, leading to the final configuration reported in Section 4. The objective was to train the model to accurately predict the Output label given the Instruction and Input code snippet.
\end{itemize}

\section{Results and Discussion}
\subsection{Baseline Performance: Un-tuned Codegen-Mono}
Prior to fine-tuning, we assessed the zero-shot performance of the base codegen-mono model on a randomly selected subset of 100 examples from our CWE dataset. Strikingly, in this baseline evaluation, the un-tuned codegen-mono model failed to detect a single CWE within any of the code snippets presented. This indicates that while pre-trained on a large corpus of code, the base model lacks the specific knowledge and instruction-following capabilities necessary for accurate CWE identification, at least in a zero-shot setting and without task-specific fine-tuning. This starkly underscores the need for targeted fine-tuning to adapt such models for specialized security tasks like CWE detection.

\begin{table}[!ht]
\renewcommand{\arraystretch}{1} 
\caption{Performance Metrics of Fine-tuned Model}
\label{tab:performance_metrics}
\centering
\begin{tabular}{lc} 
\toprule
\textbf{Metric} & \textbf{Value} \\
\midrule
Accuracy    & 99\% \\
Precision   & $\approx$ 98.08\% \\ 
Recall      & 100\% \\
F1-Score    & $\approx$ 99.04\% \\
\bottomrule
\end{tabular}
\end{table}

\subsection{Performance of Fine-tuned Codegen-Mono}
After instruction-following fine-tuning on our CWE dataset, the performance of codegen-mono underwent a dramatic transformation. We evaluated the fine-tuned model on a held-out test set comprising 100 instances, and the results demonstrate a remarkable level of accuracy in CWE detection. Table \ref{tab:performance_metrics} summarizes the key performance metrics achieved by the fine-tuned model. As evident from Table \ref{tab:performance_metrics}, the fine-tuned codegen-mono model achieved near-perfect accuracy of 99\% on the CWE detection task. This high accuracy is further reinforced by a precision of approximately 98.08\%, indicating that out of all instances identified as containing CWEs, the model was correct in the vast majority of cases. Furthermore, the model achieved a perfect recall of 100\%, signifying that it successfully detected all instances of CWEs present in the test set. The resulting F1-score of approximately 99.04\%, which harmonically balances precision and recall, confirms the overall exceptional performance of the fine-tuned model.

\subsection{Hardware performance metrics}
Performance metrics were evaluated on a desktop system equipped with a 13th Gen Intel(R) Core(TM) i5-13500 CPU (2.5 GHz, 14 Cores, 20 Threads) and 15.6 GB RAM, without GPU acceleration. Key inference timing results for the fine-tuned 350M model are summarized in Table~\ref{tab:inference_performance}.
The model performed reasonably well for real-time application on a moderate hardware.
These results represent a substantial and statistically significant improvement compared to the baseline performance of the un-tuned model. The fine-tuned codegen-mono has demonstrably acquired a strong capability for accurately identifying CWEs in Python code through instruction-following fine-tuning using our specifically curated dataset. This highlights the effectiveness of our approach and the potential of even small language models, when appropriately fine-tuned, to deliver high-performance solutions for specialized security tasks like on-premise CWE detection. Table \ref{tab:summary} summarizes the current trends among the researchers for CWE detection.

\begin{table}[!ht]
\renewcommand{\arraystretch}{1} 
\caption{CPU Inference Timing Metrics (codegen-mono 350M)}
\label{tab:inference_performance}
\centering
\begin{tabular}{lc} 
\toprule
\textbf{Metric} & \textbf{Value} \\
\midrule
Time to First Token (TTFT) & 0.253 seconds \\
Tokens per Second (TPS)    & 6.01 tokens/sec \\
Median Latency (P50)       & 0.165 seconds \\
P95 Latency                & 0.182 seconds \\
P99 Latency                & 0.239 seconds \\
\bottomrule
\end{tabular}
\end{table}

\section{Limitations and Future Work}
Our study, while demonstrating the potential of fine-tuned SLMs for CWE detection, has limitations. Its scope was confined to the MITRE Top 25 CWEs in Python, using a modest-sized dataset of synthetic snippets. This restricts known applicability and raises questions about generalization to real-world code complexity, other vulnerabilities, different languages, or vulnerabilities spanning multiple files. The model's performance on independent benchmarks and its inherent explainability also remain open questions.

Future work should directly address these limitations. Key priorities include expanding the scope to more CWEs and languages, enriching datasets with real-world examples, and improving model functionality towards localization and fix suggestions. Exploring alternative SLM architectures, advanced fine-tuning methods, and conducting rigorous comparative benchmarks against SAST tools and LLMs are also crucial next steps.

Furthermore, investigating model explainability techniques and piloting the integration into real-world development environments will be vital for assessing practical utility and fostering adoption. These efforts will help determine the true extent to which fine-tuned SLMs can serve as robust, trustworthy components in the software security toolkit.

\section{Conclusion}
In this paper, we have demonstrated the successful application of instruction-following fine-tuning to adapt a small language model, codegen-mono, for high-accuracy Common Weakness Enumeration (CWE) detection in Python code in a reasonable hardware. Our results show a remarkable transformation from a baseline model incapable of detecting CWEs to a fine-tuned model achieving near-perfect performance metrics. This research underscores the significant potential of fine-tuned Small Language Models as a practical, resource-efficient, and privacy-preserving solution for on-premise code security analysis, particularly for organizations handling confidential codebases. By providing a viable alternative to cloud-dependent LLM security tools, our work paves the way for broader adoption of advanced language model technology in security-sensitive environments, contributing to more secure software development practices and reduced vulnerability risks in critical applications. We encourage future research to build upon these findings by exploring larger and more diverse datasets, extending the approach to other programming languages, and developing robust, real-world CWE detection tools based on fine-tuned Small Language Models.

\section*{Data Availability}
The dataset can be found at \url{https://huggingface.co/datasets/floxihunter/synthetic_python_cwe}
The tuned model can be found at \url{https://huggingface.co/floxihunter/codegen-mono-CWEdetect}

\end{document}